\documentclass[aps,prb,twocolumn,superscriptaddress,showpacs,floatfix,amsmath,amssymb]{revtex4-1}
\usepackage{graphicx,xcolor}
\usepackage{mathtools}

\def\etal{{\em{et al.}}}	
\def\i{\mathbf{i}}
\def\j{\mathbf{j}}
\def\l{\mathbf{l}}
\def\I{\mathbf{I}}
\def\J{\mathbf{J}}
\def\L{\mathbf{L}}
\def\k{\mathbf{k}}
\def\K{\mathbf{K}}

\begin{document}
 
\title{Generalized Multiband Typical Medium Dynamical Cluster Approximation: Application to (Ga,Mn)N}
 
\author{Yi Zhang}
\email{zhangyiphys@gmail.com}
\affiliation{Department of Physics \& Astronomy, Louisiana State University, Baton Rouge, Louisiana 70803, USA}
\affiliation{Center for Computation \& Technology, Louisiana State University, Baton Rouge, Louisiana 70803, USA}
\author{R. Nelson}
\affiliation{Institute of Inorganic Chemistry, RWTH Aachen University, Landoltweg 1, 52056 Aachen, Germany}
\author{Elisha Siddiqui}
\affiliation{Department of Physics \& Astronomy, Louisiana State University, Baton Rouge, Louisiana 70803, USA}
\author{K.-M.\ Tam}
\affiliation{Department of Physics \& Astronomy, Louisiana State University, Baton Rouge, Louisiana 70803, USA}
\affiliation{Center for Computation \& Technology, Louisiana State University, Baton Rouge, Louisiana 70803, USA}
\author{U.\ Yu}
\affiliation{Department of Physics and Photon Science, GIST, Gwangju 61005, South Korea}
\author{T.\ Berlijn}
\affiliation{Center for Nanophase Materials Sciences, Oak Ridge National Laboratory, Oak Ridge, TN 37831, USA}
\affiliation{Computer Science and Mathematics Division, Oak Ridge National Laboratory, Oak Ridge, Tennessee 37831, USA}
\author{W.\ Ku}
\affiliation{Condensed Matter Physics and Materials Science Department, Brookhaven National Laboratory, Upton, New York 11973, USA}
\author{N. S.\ Vidhyadhiraja}
\affiliation{Theoretical Sciences Unit, Jawaharlal Nehru Centre For Advanced Scientific Research, Jakkur, Bangalore 560064, India}
\author{J.\ Moreno}
\affiliation{Department of Physics \& Astronomy, Louisiana State University, Baton Rouge, Louisiana 70803, USA}
\affiliation{Center for Computation \& Technology, Louisiana State University, Baton Rouge, Louisiana 70803, USA}
\author{M.\ Jarrell}
\affiliation{Department of Physics \& Astronomy, Louisiana State University, Baton Rouge, Louisiana 70803, USA}
\affiliation{Center for Computation \& Technology, Louisiana State University, Baton Rouge, Louisiana 70803, USA}

\begin{abstract}
We generalize the multiband typical medium dynamical cluster approximation and the formalism introduced by Blackman, Esterling and Berk so that it can deal with localization in multiband disordered systems with both diagonal and off-diagonal disorder with complicated potentials. We also introduce a new ansatz for the momentum resolved typical density of states that greatly improves the numerical stability of the method, while preserving the independence of scattering events at different frequencies.  Starting from the first-principles effective Hamiltonian, we apply this method to the diluted magnetic semiconductor Ga$_{1-x}$Mn$_x$N, and find the impurity band is completely localized for Mn concentrations $x<0.03$, while for $0.03 <x<0.10$ the impurity band has delocalized states but the chemical potential resides at or above the mobility edge.  So, the system is always insulating within the experimental compositional  limit ($x\approx 0.10$) due to Anderson localization.  However, for $0.03 <x<0.10$ hole doping could make the system metallic allowing double exchange mediated, or enhanced, ferromagnetism.  The developed method is expected to have a large impact on first-principles studies of Anderson localization.
\end{abstract}

\pacs{71.23.An,61.72.uj,72.15.Rn,71.23.-k}

\maketitle
 
\section{Introduction}
\label{sec:intro} 
Disorder is ubiquitous in materials and can drastically affect their properties, especially their 
electronic structure and transport properties. It can induce localization of electrons and lead to a 
metal-insulator transition, which is known as the Anderson localization transition~\cite{p_anderson_1958,e_abrahams_1979}. Signatures of Anderson localization are reported in many materials such as doped semi-conductors~\cite{a_richardella_10,m_winkler_2011}, polycrystalline phase-change materials~\cite{t_siegrist2011,w_zhang2012} and single crystal Li$_x$Fe$_7$Se$_8$~\cite{Ying1501283}, where disorder plays an important role in the detected metal-insulator transition.   

In order to capture Anderson localization in an effective medium theory, Dobrosavljevi\'{c} \etal, 
introduced the typical medium theory (TMT)~\cite{v_dobrosavljevic_03}
which is an extension of the coherent potential approximation (CPA).~\cite{p_soven_67,b_velicky_68}
In the CPA, the disordered lattice is replaced by an impurity embedded in an arithmetically averaged momentum-independent effective medium. While the CPA has been successful in describing some 
one-particle properties, such as the density of states (DOS) in substitutionally disordered 
alloys~\cite{p_soven_67, s_kirpatrick_70}, it fails to describe the Anderson 
localization transition.  This failure stems from the arithmetic average used to define the 
effective medium, which always favors the metallic state. In the TMT the arithmetic average 
DOS is replaced by the geometric average, or typical  DOS~\cite{v_dobrosavljevic_03,m_janssen_98,m_janssen_94,a_mirlin_94,e_crow_88},
which vanishes at the localization transition and can therefore serve as a proper order parameter for Anderson localization.

Although the TMT captures the localization transition, it underestimates the critical disorder strength, due to its local nature. A cluster extension of TMT, the typical medium dynamical cluster approximation (TMDCA) was developed recently~\cite{c_ekuma_14}.  It predicts a more accurate critical disorder strength and captures the re-entrant behavior of the mobility edge for the single band Anderson model with a box disorder potential. Generalizations of the TMDCA for multiband systems~\cite{y_zhang_15}, and systems with off-diagonal disorder~\cite{h_ter_14} using the Blackman, Esterling, and Berk (Blackman)~\cite{j_blackman_71} formalism, were also developed to study more complicated disordered systems. In this paper, we combine these methods to study a multiband model with both diagonal and off-diagonal disorder, which is suitable for any general disordered system, and we introduce and validate a new ansatz for the treatment of disorder which greatly enhances the stability and applicability of the method.  

We apply this method to one of the diluted magnetic semiconductors, (Ga,Mn)N.  Diluted magnetic semiconductors (DMS) are ideal candidates for spintronic device applications~\cite{i_zutic_04} including the design of nonvolatile computer memory~\cite{g_prinz_98a,s_dassarma_01}, electric field controlled 
magnetization~\cite{Ohno00,i_stolichnov_08,d_chiba_08} and spin-generating solar cells~\cite{i_zutic_01,b_endres_13}. The Ga$_{1-x}$Mn$_x$N based DMS has attracted great interest and is extensively studied, due to its close relationship to blue LED~\cite{h_amano_86,i_akasaki_93,s_nakamura_94} 
technology whose host compound is GaN. Mn doping induces ferromagnetism making Ga$_{1-x}$Mn$_x$N 
a good candidate for spintronic devices.  The efficiency of these devices depends on the Curie temperature (T$_c$) of the ferromagnetic order. Extensive experimental studies of the ferromagnetic order have been done on both zincblende and wurtzite Ga$_{1-x}$Mn$_x$N, leading to various values of T$_c$, including low T$_c$ around 10 K in zincblende~\cite{s_novikov_05} and wurtzite~\cite{m_overberg_01,s_stef_13,m_sawicki_12} structures, and high 
T$_c$ around room temperature in zincblende~\cite{v_chitta_04} and wurtzite~\cite{g_thaler_02,m_reed_01,t_sasaki_02} samples.

There are three predominant theoretical models proposed to understand the ferromagnetism in DMS.
First, the mean-field Zener model of Dietl has been the accepted paradigm for these systems~\cite{t_dietl_00,t_dietl_01,t_jungwirth_06,a_macdonald_05a} until relatively recently.
Here, a magnetic exchange between localized Mn moments mediated by the valence band holes drives 
the magnetism. Based on this, the T$_c$ of zincblende Ga$_{1-x}$Mn$_x$N with $x$=5\% is predicted 
to be higher than room temperature. Second, an impurity band based theory states that the 
ferromagnetism is due to a double-exchange coupling mediated by the impurity levels~\cite{k_sato_10,k_sato_04}. This theory is supported by evidence that the magnetic properties of (Ga,Mn)As are determined by the location 
of the chemical potential in this distinct impurity band brought on by Mn doping, and even by the Anderson localization of the impurity band carriers~\cite{Dobrowolska12,m_sawicki_10a,m_flatte_11,n_samarth_12}.
A direct experimental probe of the impurity band states in (Ga,Mn)As by scanning 
tunneling spectroscopy shows that the local density of states obtains a log-normal distribution 
at the verge of the localization transition~\cite{a_richardella_10} and the typical value of the distribution 
is vanishing. This could also happen in (Ga,Mn)N and the fact that DMSs can undergo an Anderson 
localization transition makes the study of DMSs more challenging,  especially if we consider the 
competition between localization and magnetism in (Ga,Mn)N, which is not well understood.
Third, and most recently, the ferromagnetic coupling in insulating systems was also interpreted in terms 
of super exchange.~\cite{s_stef_13,m_sawicki_12} In this paper, as an illustration of our formalism, we systematically study the metal-insulator transition due to  localization in the ferromagnetic phase of (Ga,Mn)N. We find that (Ga,Mn)N is always insulating within the compositional limit consistent with transport measurements.\cite{transport_GaMnN} Our results indicate that both the second and the third models are important to explain the magnetism in this material. For relatively high doping, double exchange might be more important due to the finite density of delocalized states in the impurity band, and for low concentrations, since the impurity band is completely localized, superexchange should play a more important role.

\section{Formalism}
\label{sec:formalism}

To study the effect of disorder in zincblende (Ga,Mn)N, we use the generalized spin-Fermion Hamiltonian~\cite{r_nelson_15} generated by the first-principles Effective Hamiltonian Method~\citep{t_berl_11}: $H_{eff}=H_0+\Delta$, where
\begin{equation}\label{eq:DFT1}
 H_{0}=\sum_{\mathbf{i,i'} m,m',\sigma}t_{\mathbf{ii'}}^{mm'}c_{im\sigma}^{\dagger}c_{\mathbf{i'} m'\sigma}+h.c.
\end{equation}
is the Hamiltonian of the pure GaN, and 
\begin{equation}\label{eq:DFT2}
\begin{split}
 \Delta&=\sum_{\mathbf{j}}\Delta_{\mathbf{j}}^{imp} =\sum_{\mathbf{j,i,i'},m,m',\sigma}T_{\mathbf{jii'}}^{mm'}c_{\mathbf{i}m\sigma}^{\dagger}c_{\mathbf{i'}m'\sigma}\\&
  +\sum_{\mathbf{j,i,i'},m,m',\sigma,\sigma'}J_{\mathbf{jii'}}^{mm'}c_{\mathbf{i}m\sigma}^{\dagger}
  \boldsymbol{\mathbf{\tau}}_{\sigma\sigma'}c_{\mathbf{i'}m'\sigma'}\cdot\mathbf{S_{j}}+h.c.
\end{split}
\end{equation}
contains the impurity potentials $\Delta_{\mathbf{j}}^{imp}$ induced by replacing one Ga with one Mn in the primitive unit cell $\mathbf{j}$. 
Here, $c_{\mathbf{i}m\sigma}^{+}$ and $c_{\mathbf{i}m\sigma}$ are the creation and annihilation operators of 
an electron with spin $\sigma$ at unit cell $\mathbf{i}$ in the $m$-th effective $\widetilde{N}$-$sp^3$ Wannier orbital.  $t_{\mathbf{ii'}}^{mm'}$ contains the bare orbital energy and hopping integral of the pure GaN. 
$T_{\mathbf{jii'}}^{mm'}$ and $J_{\mathbf{jii'}}^{mm'}$ are the spin-independent and spin-dependent impurity potentials, respectively.
$\mathbf{S_{j}}$ is the spin-$\frac{5}{2}$ unit-vector and $\tau_{\sigma\sigma'}$ are the elements of Pauli's matrices and $h.c.$
denotes the Hermitian conjugate. More details of the first-principles calculation can be found in Ref. ~\onlinecite{r_nelson_15}.
For impurity concentration $x$, i.e. Ga$_{1-x}$Mn$_x$N, the disorder potential is sampled independently on each cluster site with a binary probability density distribution function:
\begin{equation}
P(\Delta_{\mathbf{j}})=x\delta(\Delta_{\mathbf{j}}-\Delta_{\mathbf{j}}^{imp})+(1-x)\delta(\Delta_{\mathbf{j}}-0). 
\end{equation}
with $\Delta_\j$ the potential at site $\j$.

As pointed out in the Appendix of Ref.~\onlinecite{y_zhang_15}, the critical behavior of the typical DOS for each orbital is independent of the basis, as long as the basis is local. In the downfolding procedure, used here to obtain the Hamiltonian, a series of local rotations are used to generate a Hamiltonian which is block diagonal in a low energy window, while retaining the local nature of the orbital basis. Since the choice of the downfolding basis will not change the critical behavior of the typical DOS, i.e., the order parameter of the Anderson localization, it is possible to incorporate the first-principles calculation within TMDCA to study localization effects in real materials with disorder without further approximations.

From the leading parameters of the impurity potential listed in Table.~\ref{tab:impurity_p}, we see that the diagonal disorder potentials $T_{\mathbf{jii}}^{mm}$ and $J_{\mathbf{jii}}^{mm}$ are very strong and short-ranged, extending only up to nearest neighbors. The off-diagonal disorder potential $T_{\mathbf{jji}}^{mm'}$ and 
$J_{\mathbf{jji}}^{mm'}$ which are directly related to the hopping integrals from and to the impurity site are not weak
and can not be ignored. So to solve this Hamiltonian using effective medium theory, we need to adopt the Blackman formalism to deal with the off-diagonal disorder potentials between pairs.  In addition to the disorder potentials listed in Table.~\ref{tab:impurity_p}, we also find that the
impurity has a significant effect on the hopping integrals between sites that are different from but close to the
impurity site. We denote these hopping integrals as the non-local off-diagonal disorder potentials represented by the parameters
$T_{\mathbf{jii'}}^{mm'}$ and $J_{\mathbf{jii'}}^{mm'}$ whose leading strength are about 498 meV for $T$ and 336 meV for $J$.  
In order to consider these impurity potentials in our calculation, we slightly modify the Blackman formalism. These developments are described in Appendix~\ref{appendixB}.

\begin{table}[h]
\begin{tabular}{|c|c|c|c|c|}
\hline 
 & $T_{\j\i\i}^{mm}$ & $T_{\j\j\i}^{mm'}$ & $J_{\j\i\i}^{mm}$ & $J_{\j\j\i}^{mm'}$\tabularnewline
\hline 
$\i=\j$ & 2488 & -170 & 1752 & -633\tabularnewline
\hline 
$\i=NN(\j)$ & 406 & 885 & 449 & 800\tabularnewline
\hline 
$\i=NNN(\j)$ & 15 & 68 & $<$10 & 38\tabularnewline
\hline 
\end{tabular}
\caption{Leading parameters of the impurity potential in meV near the impurity
site $\j$. NN($\j$) and NNN($\j$) denotes nearest neighboring and next nearest
neighboring sites and $m\protect\neq m'$ from Ref. ~\onlinecite{r_nelson_15}.}
\label{tab:impurity_p}
\end{table}

We use a combination of the multiband DCA and TMDCA with a modified Blackman formalism to study Anderson localization in the first-principles effective Hamiltonian of (Ga,Mn)N.
We assume the system is already in the ferromagnetic phase, so that the local spins induced by the Mn impurities are pinned to 
some certain direction, which is set to the quantized direction of the electron spins. Then the two spin species are decoupled in the Hamiltonian and each contains four effective $\widetilde{N}$-$sp^3$ Wannier orbitals.  So for each spin species, we are dealing with a four-band Anderson model with both diagonal and 
off-diagonal disorder. We find that if we directly generalize the multiband TMDCA ansatz of Ref.~\onlinecite{y_zhang_15} to its Blackman version, we encounter severe numerical instability problems when solving the self-consistent equations. We find that
the source of the instability comes from the Hilbert transformation used to calculate the real part of the typical Green function from the typical density of states. The Hilbert transformation connects the typical Green function at all the frequencies and makes the real component of the typical Green function a functional of its imaginary part. This means that a small error at certain frequency can spread to its neighbor frequencies, which makes the calculation numerically unstable, especially for systems with multiple bands and complicated disorder potentials. This frequency mixing is also somewhat unphysical, since the scattering processes are purely elastic, and processes at different energy are independent.  The Hilbert transformation does not cause problems for simple Hamiltonians, but for complex first-principles effective Hamiltonians with multiple bands and bare gap structures, together with off-diagonal disorder, it causes numerical instabilities which interfere with the convergence of the  calculation. 

So in this paper, we introduce a new ansatz, to calculate the typical Green function directly, without invoking the Hilbert transformation. The spirit of searching for a proper ansatz is to find one that incorporates the typical value of the local density of states, which serves as the order parameter of the Anderson localization, that becomes exact when the cluster size approaches infinity, that promotes numerical stability, and that converges quickly as the cluster size increases. Since the ansatz Eq.~(\ref{eq:ansatz}) satisfies all these features, it is a proper ansatz to capture the physics of Anderson localization.  We find that the new ansatz yields an algorithm which is more numerically stable, converges quickly with cluster size and produces physical results. It is defined as:
\begin{widetext}
\begin{equation}\label{eq:ansatz}
G_{typ}^{mm'}(K,\omega)=e^{\frac{1}{N_{c}}\sum_{\i}\left\langle \ln\left(\sum_{m}\rho_{ii}^{mm}(\omega)\right)\right\rangle }
\left(\begin{array}{cc}
\left\langle \frac{G_{c,AA}^{mm'}(K,\omega)}{{\displaystyle {\textstyle {\scriptstyle \frac{1}{N_c}\sum_{\i,m}\rho_{\i\i}^{mm}(\omega)}}}}\right\rangle  & \left\langle \frac{G_{c,AB}^{mm'}(K,\omega)}{{\displaystyle {\textstyle {\scriptstyle \frac{1}{N_c}\sum_{\i,m}\rho_{\i\i}^{mm}(\omega)}}}}\right\rangle \\
\left\langle \frac{G_{c,BA}^{mm'}(K,\omega)}{{\displaystyle {\textstyle {\scriptstyle \frac{1}{N_c}\sum_{\i,m}\rho_{\i\i}^{mm}(\omega)}}}}\right\rangle  & \left\langle \frac{G_{c,BB}^{mm'}(K,\omega)}{{\displaystyle {\textstyle {\scriptstyle \frac{1}{N_c}\sum_{\i,m}\rho_{\i\i}^{mm}(\omega)}}}}\right\rangle 
\end{array}\right)
\end{equation}
with 
\begin{equation}
G_{c,\i\i}^{mm'}(\omega)=\sum_{\K}(G_{c,AA}^{mm'}(\K,\omega)
+G_{c,BB}^{mm'}(\K,\omega)+G_{c,AB}^{mm'}(\K,\omega)+G_{c,BA}^{mm'}(\K,\omega)) 
\end{equation}
\begin{equation}\label{eq:rho}
\rho_{\i\i}^{mm'}(\omega)=-\frac{1}{\pi}\mathrm{Im}[G_{c,\i\i}^{mm'}(\omega)]
\end{equation}
\end{widetext}
where $\left\langle (\cdots)\right\rangle$ represents averaging over disorder configurations, $m,m'$ denote the band indices, $i$ denotes the site index and $A$,$B$ are the component indices in the Blackman formalism, with $A$ representing the host atoms (Ga here) and $B$ representing the impurity atoms (Mn here). Our calculation is carried out on the real frequency axis, so there is no need to perform analytic continuation and the density of states can be calculated directly from the imaginary part of the Green function (see Appendix B for more details).
This ansatz consists of a prefactor $e^{\frac{1}{N_{c}}\sum_{\i}\left\langle \ln\left(\sum_{m}\rho_{\i\i}^{mm}(\omega)\right)\right\rangle }$ which is just the geometric average or typical value of the local DOS and a normalized algebraic average Green function in Blackman formalism. The prefactor can be regarded as the order parameter of the Anderson localization transition, which can be calculated as:
\begin{equation}
\rho_{typ}(\omega)=\frac{1}{N_c}\sum_{\K,m}\sum_{AA,BB,AB,BA}-\frac{1}{\pi}\mathrm{Im}G_{typ}^{mm}(\K,\omega),
\end{equation}
so Eq.~(\ref{eq:ansatz}) contains a proper order parameter. For very weak disorder, since the geometric and algebraic average are the same, the ansatz reduces to the multiband DCA. It directly calculates the typical Green function without invoking the Hilbert transformation, which makes the typical DOS for each frequency completely independent of each other. This feature is consistent 
with the elastic scattering in disorder systems and it dramatically increases the numerical stability of the calculation. This ansatz does not recover the TMT in the $N_c=1$ limit, but for large cluster size
it converges quickly and approaches the exact results as we will show below.
Since the multiband TMT is not a physical limit, we believe our ansatz does not need to recover it for cluster 
of size $N_c=1$.

\section{Results}\label{sec:results}

\begin{figure}[h!]
 \includegraphics[trim = 0mm 0mm 0mm 0mm,width=1\columnwidth,clip=true]{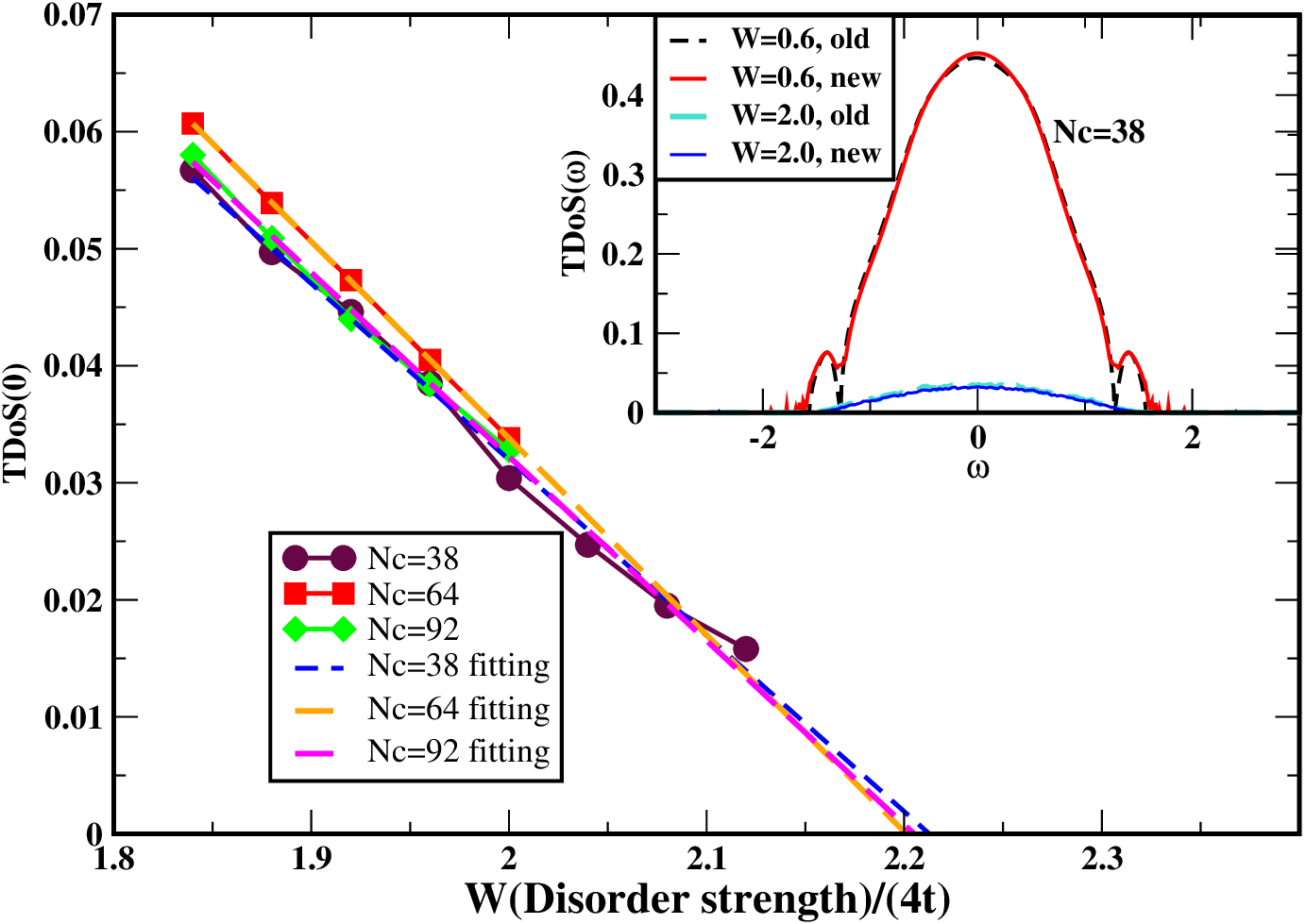}
 \caption{The typical DOS at the band center ($\omega=0$) vs.\ disorder strength with cluster size $N_c$=38, 64, and 92. The critical disorder strength is estimated by linear extrapolation and it converges to around 2.2, as it was the case for the previous ansatz.
 Inset: Plots of the typical DOS($\omega$) with $N_c$=38 for two disorder strength for the old and new ansatz. The curves practically overlap.}
 \label{fig:single}
\end{figure}

We first test the new ansatz in the single-band model Anderson Hamiltonian with nearest-neighbor hopping $t$ and onsite disorder potential $V$. As shown in Fig.~\ref{fig:single}, the new ansatz reproduces the results of the previous ansatz~\cite{c_ekuma_14} 
where a Hilbert transformation is used and captures the physics of Anderson localization for large cluster sizes.

\begin{figure}[h!]
 \includegraphics[trim = 0mm 0mm 0mm 0mm,width=1\columnwidth,clip=true]{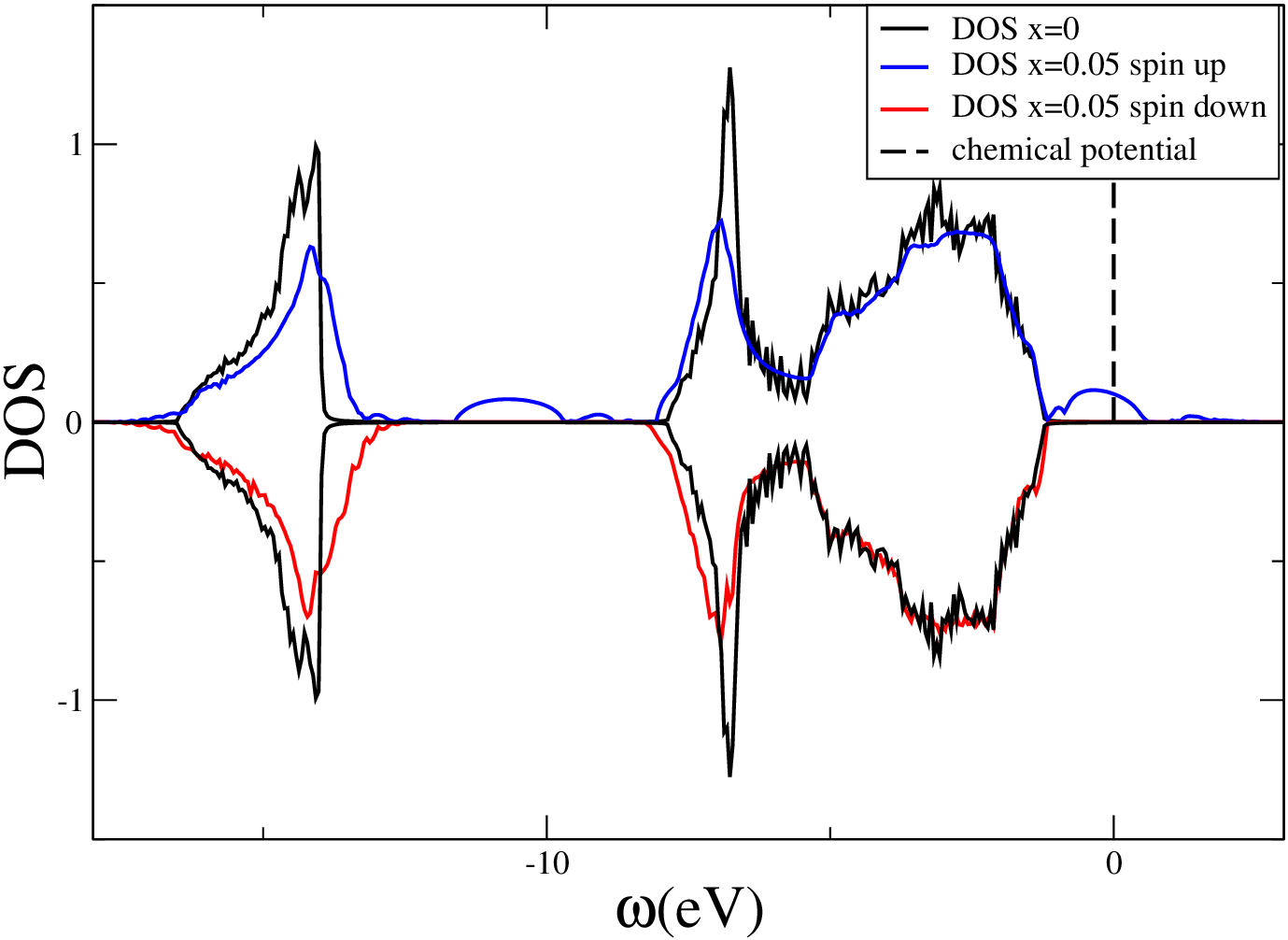}
 \caption{The DOS of Ga$_{1-x}$Mn$_x$N with $x=0.05$ 
 for both spin species (calculated with DCA and a cluster of size $N_c=32$) 
 compared with the DOS of the pure GaN (calculated directly from the downfolded first-principle Hamiltonian).The chemical potential is calculated assuming each Mn impurity contributes one hole to the system.}
 \label{fig:dos_up_down}
\end{figure}

Next we apply the multiband DCA and multiband TMDCA with the new ansatz to the Hamiltonian for Ga$_{1-x}$Mn$_x$N. In all figures we calculate the DOS of the system using DCA and the typical DOS using TMDCA. The self-consistent loop is described in Appendix~\ref{appendixB}.  It is similar to that in Ref.~\onlinecite{y_zhang_15} generalized to the Blackman  formalism as in Ref.~\onlinecite{h_ter_14}, and with some modifications to deal with non-local off-diagonal disorder.
We first calculate the DOS of both spin species for $x=0.05$. As shown in Fig.~\ref{fig:dos_up_down}, the impurity band only contains the spin up species as 
the spin down species has no impurity band around the chemical potential and is always fully filled. 
This can be understood by looking at the leading order diagonal disorder potential
in Table.~\ref{tab:impurity_p}, where $T_{\j\j\j}^{mm}=2488$ meV and $J_{\j\j\j}^{mm}=1752$ meV, 
so the disorder potential felt by the spin down channel $T_{\j\j\j}^{mm}-J_{\j\j\j}^{mm}=736$ meV
is much weaker than that felt by the spin up channel $T_{\j\j\j}^{mm}+J_{\j\j\j}^{mm}=4240$ meV.
Since we are interested in the localization of the states near the chemical potential, we will only focus on the spin up channel in our calculation below.

\begin{figure}[h!]
 \includegraphics[trim = 0mm 0mm 0mm 0mm,width=1\columnwidth,clip=true]{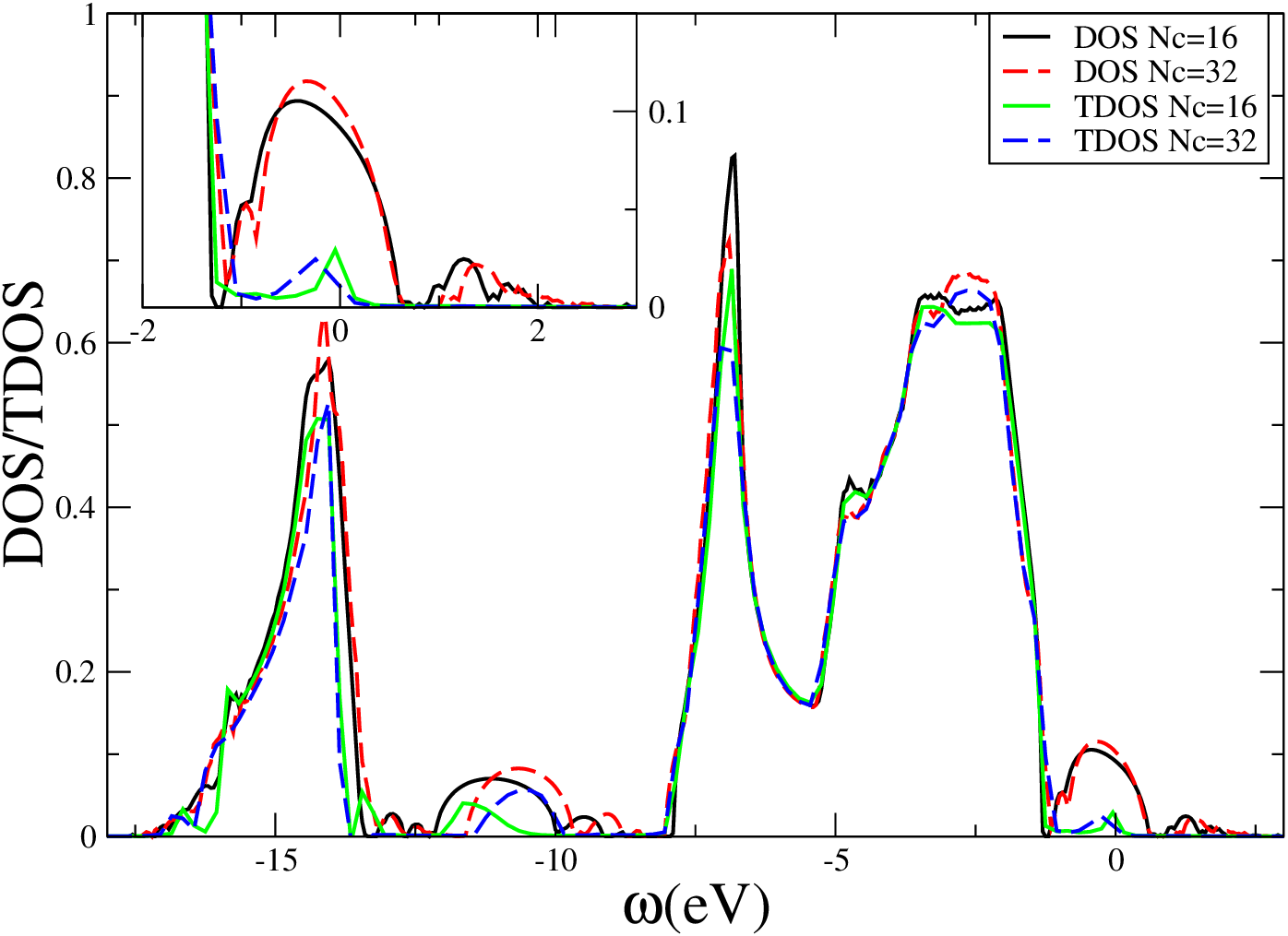}
 \caption{The spin-up DOS and TDOS of Ga$_{1-x}$Mn$_x$N with $x=0.05$ for two cluster sizes
  $N_c=16$ and $N_c=32$. The inset shows the same plot around the impurity band region.}
 \label{fig:dos_tdos_Nc}
\end{figure}

Next, we check the convergence of the DCA and the TMDCA with cluster size and find that the DOS and typical DOS calculated for clusters of size $N_c=16$ and $32$ are quite close indicating that the result is nearly converged for $N_c=32$ as shown in Fig.~\ref{fig:dos_tdos_Nc}.

\begin{figure}[h!]
 \includegraphics[trim = 0mm 0mm 0mm 0mm,width=1\columnwidth,clip=true]{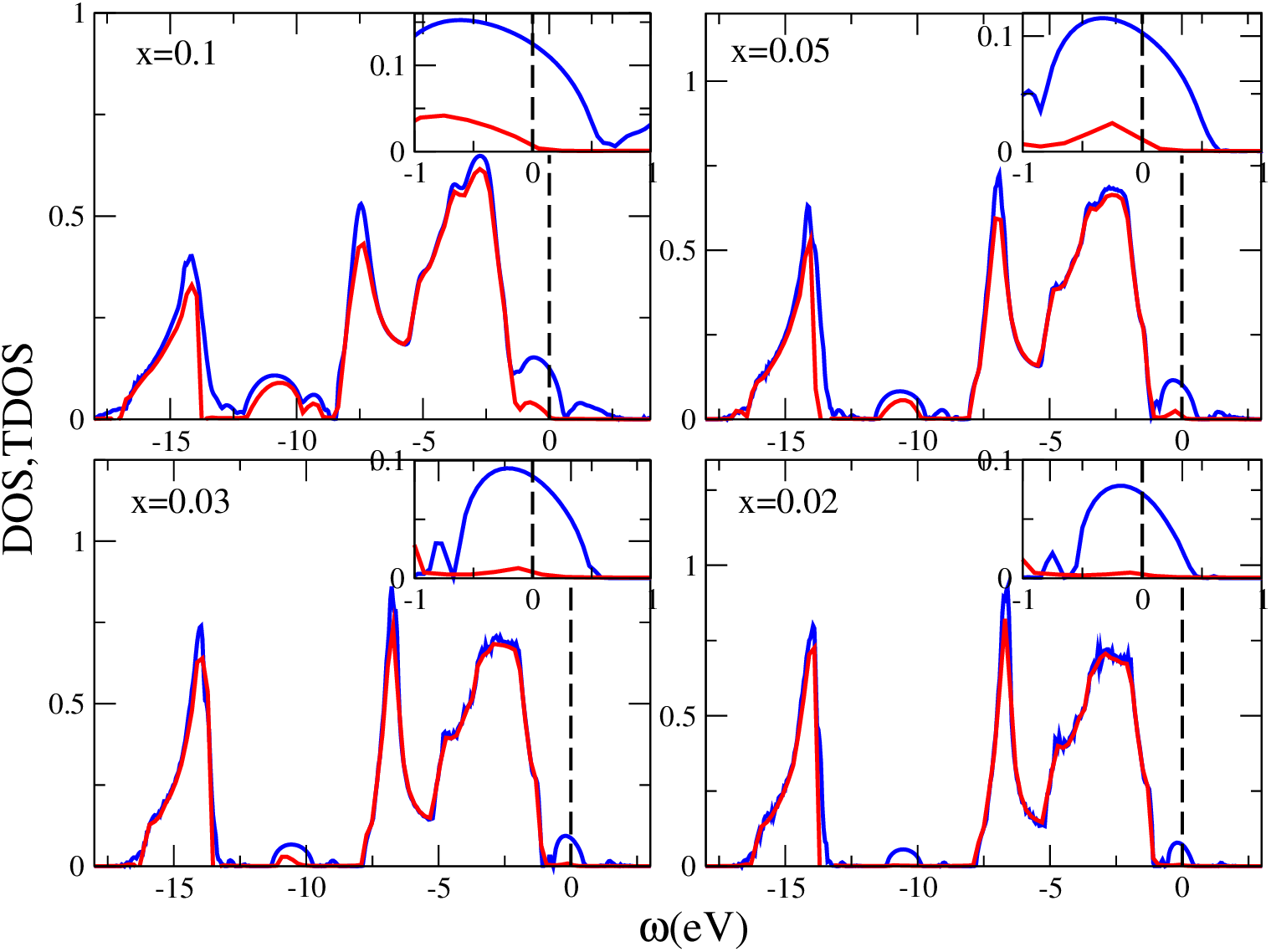}
 \caption{DOS(blue) and typical DOS(red) of Ga$_{1-x}$Mn$_x$N for various Mn concentrations: x=0.02, 0.03, 0.05, 0.1, with N$_c$=32, showing that the impurity band is completely localized for $x\le 0.03$.
 The chemical potential is set to be zero and denoted as the dash line. Inset: Zoom in of the DOS and TDOS
 around the chemical potential.}
 \label{fig:dos_tdos_mu_x}
\end{figure}

\begin{figure}[h!]
 \includegraphics[trim = 0mm 0mm 0mm 0mm,width=1\columnwidth,clip=true]{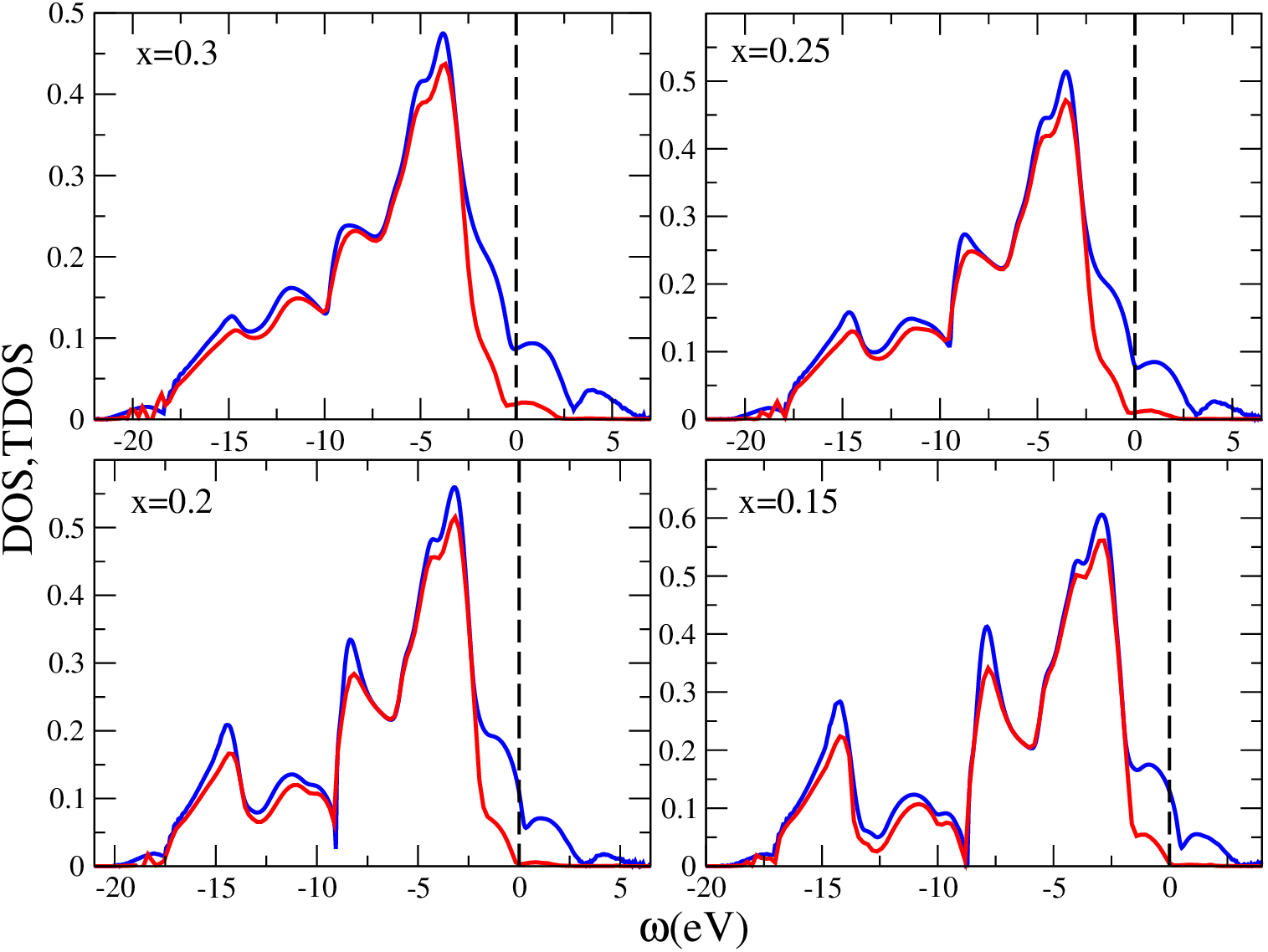}
 \caption{
 DOS(blue) and typical DOS(red) of Ga$_{1-x}$Mn$_x$N for higher Mn concentrations: x=0.15, 0.2, 0.25, 0.3, with N$_c$=32, showing a transition to the metallic state happens around $x=0.25$.
 The chemical potential is set to be zero and denoted as the dash line.}
 \label{fig:dos_tdos_mu_x2}
\end{figure}

Then to study localization, we calculate the evolution of the typical DOS and average DOS around the chemical potential as the Mn concentration $x$ increases. To determine the chemical potential, indicated by $\omega=0$, we suppose that each Mn impurity contributes one hole in the system.  Then the total electron density in the effective $\widetilde{N}$-$sp^3$ basis is $8-x$, and since spin down species are always fully filled, the density of the spin up species is $4-x$. Then we use the DCA to calculate the average DOS and determine the position of the chemical potential.  As shown in Fig.~\ref{fig:dos_tdos_mu_x}, the typical DOS of the impurity band vanishes when $x<0.03$, which means that the midgap states induced by the impurities are completely localized. As $x$ increases, the impurity band starts to hybridize with the valence band and the TDOS of the impurity band gradually increases indicating the occurrence of delocalized states in the impurity band. While the chemical potential still lies above, or on, the localization edge within our numerical accuracy, the system is still insulating. In Fig.~\ref{fig:dos_tdos_mu_x2} the chemical potential starts to cross the localization edge for $x>0.25$ indicating a transition to the metallic state. This means that Ga$_{1-x}$Mn$_x$N is always insulating due to Anderson localization within the compositional limit which is about 10\% consistent with the experimental results\cite{transport_GaMnN}. For $x<0.03$, since the impurity band is completely localized, the carriers are trapped and are not able to mediate the double exchange interactions between local spins to form long-range ferromagnetic order. In terms of our mean field theory, the host Green function is no longer polarized. The possible ferromagnetic phase may come from the super exchange as pointed out in Ref.~\onlinecite{s_stef_13,m_sawicki_12}. For $x>0.03$, a non-magnetic dopant such as Zn, or e.g., the application of a gate bias, could move the chemical potential down, leading to a metallic phase with ferromagnetism induced by double exchange as well as superexchange\cite{s_stef_13,m_sawicki_12} possibly enhancing the Curie temperature significantly.

In this work, we focus on the Anderson localization of electrons in (Ga,Mn)N, rather than the mechanism of its ferromagnetism. So we do not directly compute the magnetic properties of Ga$_{1-x}$Mn$_x$N (and, as such, neither we evaluate the superexchange). 
Instead we assume the Mn moments to be aligned ferromagnetically and investigate from first principles the Anderson localization in the impurity band. This is important because Ref.~\onlinecite{Dobrowolska12,m_sawicki_10a,m_flatte_11,n_samarth_12} proposed that the Anderson localization  suppresses the double-exchange mechanism of ferromagnetism in this class of materials.
The generalized first-principles spin-Fermion Hamiltonian 
of Ref.~\onlinecite{r_nelson_15} provides an impurity potential which has both spin-dependent and spin-independent components. The assumption of ferromagnetism implies that we are underestimating the effects of disorder, since we only include chemical but not magnetic disorder. Nevertheless, we find that for Mn concentrations less than 3\% the impurity band is fully localized and, therefore, the ferromagnetism is unlikely due to the double exchange mechanism. This supports the dominance of ferromagnetic superexchange for concentrations less than 3\%. 

The generalized first-principles spin-Fermion Hamiltonian of 
Ref.~\onlinecite{r_nelson_15} used in the current work does not incorporate superexchange processes described in Ref.~\onlinecite{s_stef_13,m_sawicki_12}. One may wonder, 
whether such high order perturbative corrections could suppress the Anderson 
localization. We have two reasons to think this is not the case. First of all, 
experimentally it is found that Ga$_{1-x}$Mn$_x$N is an insulator with variable-range hopping 
behavior\cite{transport_GaMnN}, a Hall-Mark of 
Anderson localization. Secondly, in the current model we have assumed the moments to 
be perfectly ordered which underestimates the effects of disorder, since only chemical 
disorder but not magnetic disorder is considered. So we expect the finding of the 
Anderson localization is robust against the inclusion of superexchange.
It will be interesting to study the Anderson localization in the presence of 
superexchange. To include superexchange we would need to redo the analysis 
in Ref.~\onlinecite{r_nelson_15} without removing the Mn-d charge degrees of freedom. 
However, to determine localization in such an interacting and 
disordered multi-band model is a formidable task, beyond the scope of the current manuscript.


\section{Conclusion}\label{sec:conclusion}
We combine the multiband TMDCA with the Blackman formalism to study multiband systems 
with both diagonal and off-diagonal disorder. We extend the Blackman formalism to 
describe off-diagonal disorder potentials which are not pairwise.  We also introduce a 
new TMDCA ansatz to overcome the numerical instability problem caused by the Hilbert 
transformation.  We tested our new ansatz with the single-band Anderson model, where it 
reproduces previous results for large cluster sizes. Our developed method will allow 
first-principles studies of many functional materials in which Anderson localization 
plays an important role.  We apply our new generalized multiband TMDCA ansatz to the 
diluted magnetic semiconductor Ga$_{1-x}$Mn$_x$N, using a first-principles tight-binding 
spin-fermion model, and predict that the impurity band is completely localized for a Mn 
concentration of less than 3\% and, since the chemical potential lies at or above the 
localization edge,  the system is always insulating within the compositional limit of 
10\%. This implies that ferromagnetism in Ga$_{1-x}$Mn$_x$N for x $\le$ 0.03
cannot be mediated by double exchange, which would require itinerant carriers in the 
impurity band. For larger concentrations, chemical doping or the application of a
gate bias could move the chemical potential down, leading to a metallic phase with 
ferromagnetism induced by double exchange as well as 
superexchange\cite{s_stef_13,m_sawicki_12} possibly enhancing the Curie temperature significantly.

\textit{Acknowledgments}--
We thank D. Young for useful discussion on the results.
This material is based upon work supported by the National Science Foundation under the Cooperative Agreement No. EPS-1003897 with additional support from the Louisiana Board of Regents.  Work by TB was performed at the Center for Nanophase Materials Sciences, a DOE Office of Science user facility. This manuscript has been authored by UT-Battelle, LLC under Contract No. DE-AC05-00OR22725 with the U.S. Department of Energy. WK was supported by DOE Contract No. DEAC02-98CH10886. This work used the high performance computational resources provided by the Louisiana Optical Network Initiative (http://www.loni.org), and HPC@LSU computing. The United States Government retains and the publisher, by accepting the article for publication, acknowledges that the United States Government retains a non-exclusive, paid-up, irrevocable, world-wide license to publish or reproduce the published form of this manuscript, or allow others to do so, for United States Government purposes. The Department of Energy will provide public access to these results of federally sponsored research in accordance with the DOE Public Access Plan (http://energy.gov/downloads/doepublic-access-plan). 

\appendix

\section{Extraction of the impurity potential from the first-principles effective Hamiltonian and its incorporation into the Blackman formalism}
\label{appendixA}
We start from the first-principles effective Hamiltonian of Eq.~(\ref{eq:DFT1}) and (\ref{eq:DFT2}), where $\Delta_j$ contains the impurity
potential induced by the impurity located at site $\mathbf{j}$. Since for each impurity, the induced impurity potential in neighboring sites
has the same form, we can rewrite the parameters in Eq.~(\ref{eq:DFT2}) as:
\begin{equation}
T_{\j\i\i'}^{mm'}= T_{\i-\j,\i'-\j}^{mm'}
\end{equation}
\begin{equation}
J_{\j\i\i'}^{mm'}= J_{\i-\j,\i'-\j}^{mm'}
\end{equation}
Here, since the spin-independent and spin-dependent parameters have similar structures, we only show the transformation for
the spin-independent parameter. The spin-dependent component can be inferred by analogy.

To investigate the structure of the impurity potential, we first look at the terms induced by a single impurity located at the origin $\Delta_{0}$, and further split it into three parts:
\begin{equation}
\begin{split}
\Delta_{0}&=\sum_{\i,\i',m,m',\sigma}T_{\i\i'}^{mm'}c_{\i m\sigma}^{+}c_{\i' m'\sigma}\\&
=\sum_{\i,m,m',\sigma}T_{\i\i}^{mm'}c_{\i m\sigma}^{+}c_{\i m'\sigma}
+\sum_{\i\ne0,m,m',\sigma}T_{0\i}^{mm'}c_{0m\sigma}^{+}c_{\i m'\sigma}\\&
+\sum_{\i,\i'\ne0,\i\ne \i',m,m',\sigma}T_{\i\i'}^{mm'}c_{\i m\sigma}^{+}c_{\i' m'\sigma}+h.c.
\end{split}
\end{equation}
The first term is diagonal disorder which in general extends to a finite region from the origin. The second term is the off-diagonal disorder associated with hopping between the impurity site and a host site. The disorder induced by this term can be properly described in the Blackman formalism.  The last term is the off-diagonal disorder associated with the hopping between two host sites.  The disorder caused by this term can not be described properly in the original Blackman formalism, so a slight
modification is made to include these terms in our calculation. 

To extend the Blackman formalism we first write $H_{eff}$ for a specific disorder configuration, with impurities labeled by $\l$,
\begin{equation}
\begin{split}
H_{eff}&=H_{0}+\sum_{\l}\Delta_{\l}=\sum_{\i,m,m',\sigma}\epsilon_{\i\sigma}^{mm'}c_{\i m\sigma}^{+}c_{\i m'\sigma}\\&
+\sum_{\i \ne \j,m,m',\sigma}W_{\i,\j,\sigma}^{mm'}c_{\i m\sigma}^{+}c_{\j m'\sigma}
\end{split}
\end{equation}
where, 
\begin{equation}\label{eq:dd}
\epsilon_{\i\sigma}^{mm'}=t_{\i\i\sigma}^{mm'}+\sum_{\l}T_{\l\i\i}^{mm'},
\end{equation}
\begin{equation}
\begin{split}
W_{\i,\j,\sigma}^{mm'}&=t_{\i\j}^{mm'}+\sum_{\l}T_{\l\i\j}^{mm'}\\&
=t_{\i\j}^{mm'}+\sum_{\l=\i,or,\j}T_{\l\i\j}^{mm'}+\sum_{\l\ne \i,\l\ne \j}T_{\l\i\j}^{mm'}.
\end{split}
\end{equation}
Here, the first term is independent of the disorder configuration.  The third term depends on the disorder configuration but is independent of the chemical occupation of sites $\i$ and $\j$.  The second term only depends on the chemical occupation of sites $\i$ and $\j$.  If we denote the site as A if it is occupied by the host atom and B if it is occupied by the impurity atom, then we can see there are only four possible values for the second term:

\begin{equation}
\sum_{\l=\i,or,\j}T_{\l\i\j}^{mm'}=\begin{cases}
\begin{array}{c}
0,\ \ \ \ if\ \i \in A,\ \j \in A\\
T_{\j\i\j}^{mm'},\ \ \ \ if\ \i \in A,\ \j \in B\\
T_{\i\i\j}^{mm'},\ \ \ \ if\ \i \in B,\ \j \in A\\
T_{\j\i\j}^{mm'}+T_{\i\i\j}^{mm'},\ \ \ \ if\ \i \in B,\ \j \in B,
\end{array}\end{cases}
\end{equation}
so in the Blackman formalism, the hopping term $W_{\i,\j,\sigma}^{mm'}$can
be written as 2 by 2 block matrix:
\begin{equation}
\begin{split}
\underline{W}_{\i,\j,\sigma}^{mm'}&=t_{\i\j}^{mm'}\left[\begin{array}{cc}
1 & 1\\
1 & 1
\end{array}\right]+\left[\begin{array}{cc}
0 & T_{\j\i\j}^{mm'}\\
T_{\i\i\j}^{mm'} & T_{\j\i\j}^{mm'}+T_{\i\i\j}^{mm'}
\end{array}\right]\\&+\sum_{\l\ne \i,\l\ne \j}T_{\l\i\j}^{mm'}\left[\begin{array}{cc}
1 & 1\\
1 & 1
\end{array}\right].
\end{split}
\end{equation}
Here, we use underscore to denote the 2 by 2 matrix in Blackman formalism and we use overbar to denote the quantities
that are coarse-grained in the cluster.
We can see that the first two terms are configuration independent
and translationally invariant in the Blackman formalism, because
\begin{equation}
T_{\j\i\j}^{mm'}=T_{\i-\j,0}^{mm'}
\end{equation}
\begin{equation}
T_{\i\i\j}^{mm'}=T_{0,\j-\i}^{mm'},
\end{equation}
so we can let 
\begin{equation}
\underline{W}_{\i,\j,\sigma}^{1,mm'}=\left[\begin{array}{cc}
t_{\i\j}^{mm'} & t_{\i\j}^{mm'}+T_{\i-\j,0}^{mm'}\\
t_{\i\j}^{mm'}+T_{0,\j-\i}^{mm'} & t_{\i\j}^{mm'}+T_{\i-\j,0}^{mm'}+T_{0,\j-\i}^{mm'}
\end{array}\right],
\end{equation}
\begin{equation}
\underline{W}_{\i,\j,\sigma}^{2,mm'}=\sum_{\l\ne \i,\l\ne \j}T_{\l\i\j}^{mm'}\left[\begin{array}{cc}
1 & 1\\
1 & 1
\end{array}\right]=\sum_{\l\ne \i,\l\ne \j}T_{\i-\l,\j-\l}^{mm'}\left[\begin{array}{cc}
1 & 1\\
1 & 1
\end{array}\right],
\end{equation}
so that 
\begin{equation}
\underline{W}_{\i,\j,\sigma}^{mm'}=\underline{W}_{\i,\j,\sigma}^{1,mm'}+\underline{W}_{\i,\j,\sigma}^{2,mm'}.
\end{equation}

Then, $\underline{W}_{\i,\j,\sigma}^{mm'}$ is coarse-grained in the DCA cluster with periodic boundary conditions to obtain the cluster parameters $\underline{\overline{W}}_{\I,\J,\sigma}^{mm'}$
which are used for the DCA and TMDCA calculations in the Blackman formalism.

Here, since $\underline{W}_{\i,\j,\sigma}^{1,mm'}$ is translationally invariant in the Blackman formalism, it can be coarse-grained easily in the same manner as the regular kinetic energy terms:
\begin{eqnarray}
 \underline{W}_{\k,\sigma}^{1,mm'}=\sum_{\i}\underline{W}_{\i,\j,\sigma}^{1,mm'}
 e^{i\k\cdot(\mathbf{r}_{i}-\mathbf{r}_{j})},
 \\
 \underline{\overline{W}}_{\K,\sigma}^{1,mm'}=\frac{N_c}{N}\sum_{\k}\underline{W}_{\K+\k,\sigma}^{1,mm'},
 \\
 \underline{\overline{W}}_{\I,\J,\sigma}^{1,mm'}=\frac{1}{N_{c}}\sum_{\K}\underline{\overline{W}}_{\K,\sigma}^{1,mm'}
 e^{-i\K\cdot(\mathbf{r}_{I}-\mathbf{r}_{J})}.
\end{eqnarray}
But $\underline{W}_{i,j,\sigma}^{2,mm'}$ still depends on the disorder configuration, 
and is not translationally invariant, so it needs to be coarse-grained differently. We carry out the the coarse-graining according to the following procedure: 
\begin{equation}
\underline{W}_{\k,\k',\sigma}^{2,mm'}=\sum_{\i,\j}\underline{W}_{\i,\j,\sigma}^{2,mm'}
e^{i(\k\cdot\mathbf{r}_{i}-\k'\cdot\mathbf{r}_{j})},
\end{equation}
\begin{equation}
\underline{\overline{W}}_{\K,\K',\sigma}^{2,mm'}=(\frac{N_{c}}{N})^{2}\sum_{\k,\k'}
\underline{W}_{\K+\k,\K'+\k',\sigma}^{2,mm'},
\end{equation}
\begin{equation}
\underline{\overline{W}}_{\I,\J,\sigma}^{2,mm'}=(\frac{1}{N_{c}})^{2}\sum_{\K,\K'}
\underline{W}_{\K,\K',\sigma}^{2,mm'}e^{-i(\K\cdot\mathbf{R}_{I}-\K'\cdot\mathbf{R}_{J})}.
\end{equation}
The diagonal disorder component from Eq.~(\ref{eq:dd}) includes also an extended contribution, 
$T_{\l\i\i}^{mm'}=T_{\i-\l,\i-\l}^{mm'}$, which needs to be coarsed grain. We implement
the following procedure:
\begin{equation}
T_{\k}^{mm'}=\sum_{\i}T_{\i\i}^{mm'}e^{i\k\cdot\mathbf{r}_{i}},
\end{equation}
\begin{equation}
\overline{T}_{\K}^{mm'}=\frac{N_{c}}{N}\sum_{\k}T_{\K+\k}^{mm'},
\end{equation}
\begin{equation}
\overline{T}_{\I\I}^{mm'}=\frac{1}{N_{c}}\sum_{\K}T_{\K}^{mm'}e^{-i\K\cdot\mathbf{R}_{I}} \,.
\end{equation}
Then the coarse-grained version of Eq.~(\ref{eq:dd}) is just
\begin{equation}
\begin{split}
\overline{\epsilon}_{\I\sigma}^{mm'}&=t_{\I\I\sigma}^{mm'}+\sum_{\L}\overline{T}_{\I-\L,\I-\L}^{mm'}
\\&
=\epsilon_{0\sigma}^{mm'}+\overline{V}_{\I}^{mm'},
\end{split}
\end{equation}
where
\begin{equation}
\overline{V}_{\I}^{mm'}=\sum_{\L}\overline{T}_{\I-\L,\I-\L}^{mm'}
\end{equation}
is the diagonal disorder potential in the cluster. Since 
$t_{\I\I\sigma}^{mm'}$ is local and translationally invariant, 
it is not modified by coarse graining, so we set it to $\epsilon_{0\sigma}^{mm'}$.
For the spin-dependent part, the same procedure can be carried out completely by analogy.

\section{Self-consistency loop of the multiband DCA/TMDCA in the Blackman formalism}
\label{appendixB}
From the procedure above, we get the parameters needed for the DCA/TMDCA calculation, which are $\overline{\epsilon}_{\I\sigma}^{mm'}=\epsilon_{0\sigma}^{mm'}+\overline{V}_{\I,\sigma}^{mm'}$
for the diagonal component and $\underline{W}_{I,J,\sigma}^{1,mm'}$ and $\underline{W}_{I,J,\sigma}^{2,mm'}$
for the off-diagonal component of the disorder potential. The self-consistent loop can be described as below.

We first introduce the effective medium hybridization function $\underline{\Delta}(\K,\omega)$, which is a matrix in the orbital and spin basis with each component a 2 $\times$ 2 matrix in the Blackman formalism,
\begin{equation}
\underline{\Delta}_{\sigma}^{m,m'}(\K,\omega)=
\left[\begin{array}{cc}
\Delta_{\sigma,AA}^{m,m'}(K,\omega) & \Delta_{\sigma,AB}^{m,m'}(K,\omega)\\
\Delta_{\sigma,BA}^{m,m'}(K,\omega) & \Delta_{\sigma,BB}^{m,m'}(K,\omega)
\end{array}\right].
\end{equation}
Then we stochastically sample the disorder configurations, and for each configuration, we calculate the cluster Green function,
\begin{equation}
G_{\I,\J}=(\omega\mathbb{I}_r\mathbb{I}_{os}-(\epsilon_0-\overline{V})\mathbb{I}_r
-\overline{W}^{1}-\overline{W}^{2}-\Delta)_{\I,\J}^{-1},
\end{equation}
where $\mathbb{I}_r$ and $\mathbb{I}_{os}$ are the identity matrices in the cluster real space 
and the spin orbital basis respectively, with matrix elements $\delta_{\mathbf{RR'}}$ and $\delta_{mm'}^{\sigma\sigma'}$. All the quantities in the bracket are 
matrices in spin orbital and cluster real space with:
\begin{equation}
\Delta_{IJ,\sigma}^{m,m'}=FT[\Delta_{\sigma,\alpha\beta}^{m,m'}(\K,\omega)],\ \ \ \ if\ \I\in\alpha,\ \J\in\beta,
\end{equation}
\begin{equation}
\overline{W}_{\I\J,\sigma}^{1(2),m,m'}=\overline{W}_{\I\J,\sigma,\alpha\beta}^{1(2),m,m'},\ \ \ \ if\ \I\in\alpha,\ \J\in\beta,
\end{equation}
where $\alpha$ and $\beta$ denote the A,B components and $FT$ denotes the Fourier transformation.

In the next step, we perform a disorder averaging and reexpand the averaged Green function into a $2N_{c}\times2N_{c}$ block matrix, with each component a matrix in the spin and orbital bases,
\begin{equation}
\underline{G}_{c}(\omega)_{\I\J}=\left[\begin{array}{cc}
\left\langle G_{c,AA}(\omega)\right\rangle_{\I\J} & \left\langle G_{c,AB}(\omega)\right\rangle_{\I\J}\\
\left\langle G_{c,BA}(\omega)\right\rangle_{\I\J} & \left\langle G_{c,BB}(\omega)\right\rangle_{\I\J}
\end{array}\right],
\end{equation}
with
\begin{equation}
(G_{c,\alpha\beta})_{\I\J}=(G_{c})_{\I\J},\ \ \ \ if\ \I\in A,\ \J\in B.
\end{equation}

After the disorder average, the translational symmetry is restored, and we then Fourier transform it to $\K$ space and construct the $\K$-dependent cluster Green function,
\begin{equation}
\underline{G}_{c}(K,\omega)=\left[\begin{array}{cc}
\left\langle G_{c,AA}(K,\omega)\right\rangle  & \left\langle G_{c,AB}(K,\omega)\right\rangle \\
\left\langle G_{c,BA}(K,\omega)\right\rangle  & \left\langle G_{c,BB}(K,\omega)\right\rangle 
\end{array}\right].
\end{equation}
For the DCA we use this cluster Green function for the remaining calculation, and for the TMDCA, we use the ansatz described in Eq.~(\ref{eq:ansatz}) to construct the typical Green function 
$\underline{G}_{typ}(\K,\omega)$ to proceed. Note that for TMDCA the disorder average is done on
the quantity $\displaystyle\frac{G_{c}}{\frac{1}{N_c}\sum_{\i,m}\rho_{\i\i}^{mm}}$.

Once the cluster problem is solved, we calculate the coarse-grained Green function as,
\begin{equation}
\begin{split}
\underline{\overline{G}}(\K,\omega)&=\left[\begin{array}{cc}
\overline{G}_{AA}(\K,\omega) & \overline{G}_{AB}(\K,\omega)\\
\overline{G}_{BA}(\K,\omega) & \overline{G}_{BB}(\K,\omega)
\end{array}\right]\\&
=\frac{N_{c}}{N}\sum_{\k}(\underline{G}_{c(typ)}(\K,\omega)^{-1}
+\underline{\Delta}(\K,\omega)+\underline{\overline{W}}_{\K}-\underline{W}_{\k})^{-1}.
\end{split}
\end{equation}

The DCA (TMDCA) self-consistency condition requires that the disorder averaged cluster Green function equals the coarse-grained lattice Green function
\begin{equation}
 \underline{G}_{c(typ)}(\K,\omega) = \underline{\overline{G}}(\K,\omega).
\end{equation}

Then, we close our self-consistency loop by updating the hybridization function matrix using linear mixing
\begin{equation}
 \underline{\Delta}_n(\K,\omega) = \underline{\Delta}_o(\K,\omega) + 
 \xi [\underline{G}_c^{-1}(\K,\omega) - \underline{\overline{G}}^{-1}(\K,\omega)], 
\end{equation}
where the subscript $``o"$ and $``n"$ denote old and new respectively, and $\xi$ is a linear mixing factor $0<\xi<1$. 
The procedure above is repeated until
the hybridization function matrix converges to the desirable accuracy
$\underline{\Delta}_n(\K,\omega) = \underline{\Delta}_o(\K,\omega)$.

\bibliography{TMDCA_GaMnN}

\end{document}